# Persistent photogenerated state attained by femtosecond laser irradiation of thin $T_d$-MoTe$_2$

Meixin Cheng[1], Shazhou Zhong[2], Nicolas Rivas[1,♦], Tina Dekker[2], Ariel Alcides Petruk[1], Patrick Gicala[1], Kostyantyn Pichugin[1], Fangchu Chen[2], Xuan Luo[3], Yuping Sun[3,4,5], Adam W. Tsen[2], Germán Sciaini[1,*]

[♦]Present address: Nuclear Engineering Group, McMaster University, Hamilton, Ontario L8S 4K1, Canada.

[*]Correspondence: gsciaini@uwaterloo.ca

[1]The Ultrafast Electron Imaging Lab, Department of Chemistry, and Waterloo Institute for Nanotechnology, University of Waterloo, Waterloo, Ontario N2L 3G1, Canada.

[2]Institute for Quantum Computing, Department of Physics and Astronomy, Department of Electrical and Computer Engineering, and Department of Chemistry, University of Waterloo, Waterloo, Ontario N2L 3G1, Canada.

[3]Key Laboratory of Materials Physics, Institute of Solid-State Physics, HFIPS, Chinese Academy of Sciences, Hefei 230031, China.

[4]High Magnetic Field Laboratory, Chinese Academy of Sciences, HFIPS, Hefei 230031, China.

[5]Collaborative Innovation Center of Advanced Microstructures, Nanjing University, Nanjing 210093, China.

**Laser excitation has emerged as a means to expose hidden states of matter and promote phase transitions on demand. Such laser induced transformations are often rendered possible owing to the delivery of spatially and/or temporally manipulated light, carrying energy quanta well above the thermal background. Here, we report time-resolved broadband femtosecond (fs) transient absorption measurements on thin flakes of Weyl semimetal candidate $T_d$-MoTe$_2$ subjected to various levels and schemes of fs-photoexcitation. Our results reveal that impulsive fs-laser irradiation alters the interlayer behavior of the low temperature $T_d$ phase as evidenced by the persistent disappearance of its characteristic coherent $^1A_1 \approx 13$ cm$^{-1}$ shear phonon mode. We found that this structural transformation withstands thermal annealing up to 500 K, although it can be reverted to the 1$T'$ phase by fs-laser treatment at room temperature. Our work opens the door to reversible optical control of topological properties.**

**Keywords:** Laser treatment, 2D materials, femtosecond spectroscopy, femtosecond transient absorption, transition metal dichalcogenides, phonon dynamics, Weyl semimetal, MoTe$_2$.

The advent of intense laser pulses[1] has made possible the generation of far-from-equilibrium states that can quickly evolve into exotic but relatively short-lived transient phases, and with that the birth of the so-called field of 'photoinduced phase transitions (PIPTs)'[2]. Recently, a persistent low temperature light-induced metastable state has been discovered in the two-dimensional transition metal dichalcogenide (2D-TMDC) 1$T$-TaS$_2$[3]; a prototypical charge density wave (CDW) system that displays several first-order thermal phase transformations among different CDW-textured phases and related PIPTs[3–6]. On the other hand, MoTe$_2$ is one of the few TMDCs that presents different polytypes with phase changes that can be further controlled by temperature[7], alloying[8], strain[9], electrostatic gating[10], and dimensionality[11]. Distortion of in-plane bonds of the conventional 1$T$ structure gives rise to the enlarged centrosymmetric unit cell of the 1$T$' phase that is stable at room temperature. Here, Mo atoms are surrounded by Te atoms in a distorted octahedral configuration, slightly shifted from the centre and with a $c$-axis inclination angle ($\alpha$) of 93.92° with respect to the plane of the layer (Fig. 1a). Cooling below a critical temperature of $T_c \sim 250$ K prompts the shear displacement of adjacent layers to transform the bulk 1$T$'-MoTe$_2$ crystalline phase into the $T_d$-MoTe$_2$ Weyl semimetal candidate state[12–14]. The latter exhibits a non-centrosymmetric unit cell with $\alpha = 90°$ (Fig. 1a). Such breaking of inversion symmetry causes the appearance of a low-frequency Raman-active $^1A_1 \sim 13$ cm$^{-1}$ interlayer shear mode (Fig. 1b). The latter has been exploited as a parameter to follow the aforementioned first-order thermal phase change[11,15,16] as well as possible PIPTs[17–20].

Furthermore, the use of femtosecond electron diffraction (FED)[21,22] has recently revealed that photoexcitation of the $T_d$-MoTe$_2$ state, in the visible and mid-infrared spectral regions, leads to an interlayer shear displacement of only few picometers (pm)[19], i.e., a much smaller magnitude than the actual displacement necessary for the complete transition to the 1$T$' phase $\approx 19$ pm[19]. This finding is also consistent with the partial interlayer shear motion of ≈ 8 pm observed via relativistic FED in the analogous $T_d$-WTe$_2$ compound following intense terahertz (THz) excitation[23]. It is now understood[20] that the necessary interlayer displacement that connects the $T_d$ and $1T'$ states (see Fig. 1a) is unlikely to proceed on the sub-picosecond (sub-ps) timescale[17,18]. The rather simple picture of two adjacent rigid layers sliding freely in opposite directions does not account for the long-range collective response that must take place for this $T_d \rightarrow 1T'$ phase transformation to occur on such coherent fashion. Local variations in activation barriers, light induced effects[24,25], the degree of electronic delocalization, and the many-body nuclear rearrangements needed for the $T_d$ state to map the configurational space of the $1T'$ phase, are some of the phenomena that should be accounted for. Note that all well-known ultrafast PIPTs to date[4–6,26–34] involve rather localized atomic motions that can be described within the frame of a fixed unit cell[33,34] or supercell[4], usually the one that belongs to the low temperature ground state.

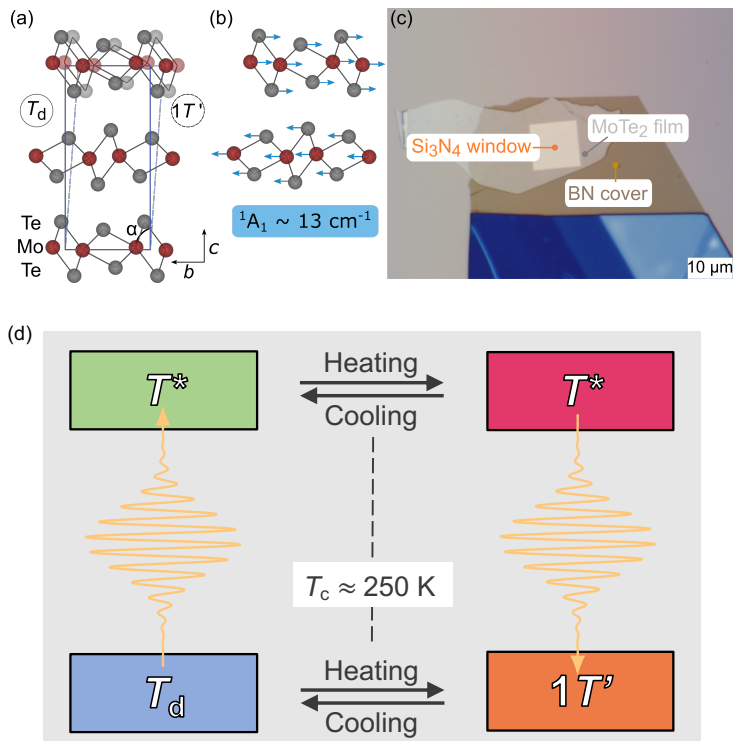

**Fig. 1:** (**a**) Unit cells of both $1T'$- and $T_d$-phases showing the change of *c*-axis tilt angle of $\Delta\alpha \approx 4°$[15]. (**b**) Illustration of the interlayer $^1A_1 \approx 13$ cm$^{-1}$ shear phonon mode. (**c**) Microscope image of a 40 nm thick $1T'$-MoTe$_2$ film obtained by mechanical exfoliation. The 50-nm thick 10 μm x 10 μm silicon nitride (Si$_3$N$_4$) square window (light brown), the $1T'$-MoTe$_2$ flake (silver), and the boron nitride (BN) protective film (brown) are clearly discernable. (**d**) Simple scheme depicting the observed fs-laser irradiation processes leading to the formation of the $T^*$ state via irradiation of the $T_d$ phase at low temperature, and its conversion to $1T'$ by irradiation at room temperature.

In this work we investigate the effects of intense femtosecond (fs) laser photoexcitation on ≈ 30 nm to 40-nm thick flakes of the Weyl semimetal candidate $T_d$-MoTe$_2$[12–14]. We will show persistent fs-laser induced changes affecting the dynamical behavior of the coherent $^1A_1$ interlayer shear phonon, and therefore the interlayer structure of $T_d$-MoTe$_2$ phase as monitored by time-resolved broadband fs transient absorption (tr-bb-TA) spectroscopy. This laser driven transformation is assigned to the formation a long-live photoinduced state, herein referred to as $T^*$.

Figure 1c exhibits a typical $1T'$-MoTe$_2$ flake that has been transferred onto a transparent 10 um x 10 um, 50-nm thick Si$_3$N$_4$ window produced via nanofabrication in house[35], and capped with h-BN for protection against ambient moisture. Figure 1d depicts the main findings in a simple scheme that links the pristine or thermodynamically stable $T_d$ and $1T'$ phases with the photogenerated $T^*$ state, illustrating the observed fs-laser induced transformations that are depicted by laser pulse arrows.

Figure 2 presents some characteristic tr-bb-TA results obtained for a 40-nm thick $T_d$-MoTe$_2$ flake at $T$ = 74 K. We observed that samples behave reversibly below incident pump fluences ($F$) ≈ 1 mJ·cm$^{-2}$. Therefore, experiments performed below this fluence threshold served to monitor the dynamical behavior of the coherent $^1A_1$ shear phonon following impulsive laser irradiation treatments. It is worth to mention that the response to impulsive photoexcitation of thin flakes was found to differ from that of a bulk $T_d$-MoTe$_2$ crystal[20], which showed structural reversibility up to at least $F$ ≈ 5 mJ cm$^{-2}$. We restricted the fittings of time traces to time delays $t \geq$ +2 ps (e.g., Fig. 2b) to enable the automated removal of the electronic background and the generation of residuals, $R_E$ (e.g., inset in Fig. 2b), from which we could extract the phonon dynamics[20].

Figure 2c displays the Fourier power spectrum obtained via the fast Fourier transform (FFT) of $R_E$ across the complete probed photon energy ($E_{probe}$) window. The Fourier amplitudes ($\mathcal{A}$) of the $^1A_1 \approx 13$ cm$^{-1}$, $^2A_1 \approx 77$ cm$^{-1}$ and $^5A_2 \approx 164$ cm$^{-1}$ Raman active modes of the $T_d$-phase and their dependencies with $E_{probe}$ are visible. The detection of higher frequency modes is currently limited by the temporal response function of our pump-probe setup $\approx 100$ fs. Figure 2d shows the Fourier spectra obtained after averaging along $E_{probe}$ and as a function of the sample temperature. The spectrum at 298 K reveals that the flake has transitioned to the $1T'$ phase as evidenced by the disappearance of $\mathcal{A}(^1A_1)$. Hence, the other two phonons with frequencies of $\approx 77$ cm$^{-1}$ and $\approx 164$ cm$^{-1}$ that are present at 298 K correspond to the $^1A_g$ and the $^8A_g$ Raman active modes of the high-temperature $1T'$ state.

Figure 3 displays the results obtained for a 32-nm thick flake at a base temperature of 220 K (just below $T_c \approx 250$ K). In this case, tr-bb-TA measurements were performed at different $F$ as specified chronologically from top to bottom in Table 1. As can be observed in Fig. 3b, $\mathcal{A}(^1A_1)$ decreases as the $F$ increases for values of $F$ above 1 mJ·cm$^{-2}$, reaching the background level at $F \approx 2$ mJ·cm$^{-2}$. After this point we assume the complete transformation of $T_d$ into $T^*$. On the other hand, no substantial changes are found in the fluence-normalized and time-averaged bb-TA spectra shown in Fig. 3a, which indicate that the material's response scales approximately linearly with $F$[20] and $T^*$ presents a similar TA spectrum to $T_d$. This latter observation is expected since the linear reflectivity spectra of the $T_d$ and $1T'$ phases are very similar in this $E_{probe}$ region that is dominated by interband transitions[36]. Note that the decrease of $\mathcal{A}(^1A_1)$ with increasing $F$ does not arise from sample damage but a structural modification that can be reverted by laser irradiation, *vide infra*. We have found that the fluence thresholds for damaging the films due to the laser's peak power vary from flake to flake and are usually in the range of $F_{th} \approx 5$ mJ·cm$^{-2}$ to 7 mJ·cm$^{-2}$. Sample damage is characterized by permanent changes in the bb-TA spectrum as well as in the Fourier amplitudes of the ≈ 77 cm$^{-1}$ and ≈ 164 cm$^{-1}$ phonon modes (see Supporting Information V, test 3).

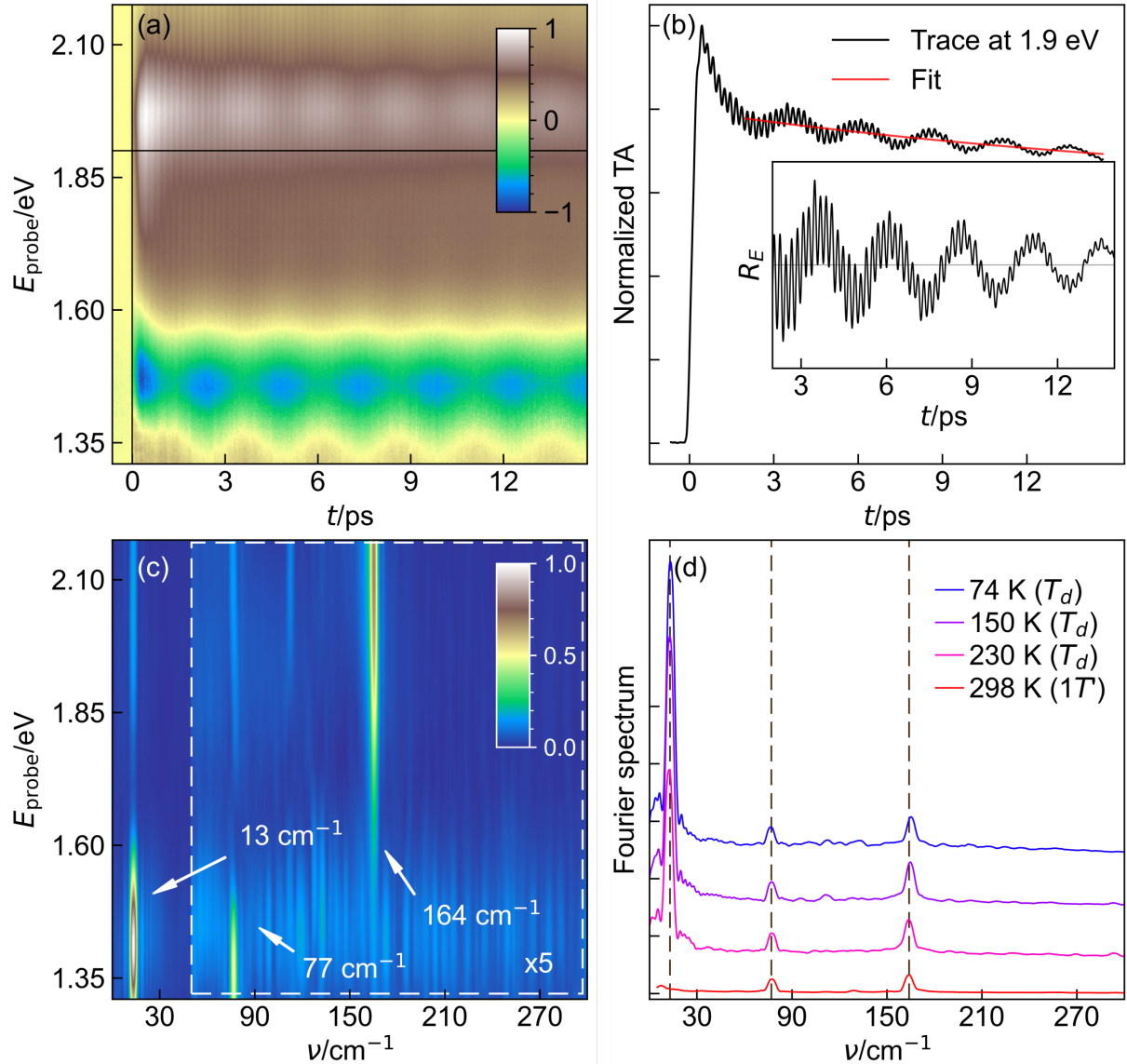

**Fig. 2: Characteristic tr-bb-TA results obtained for 40-nm thick MoTe$_2$ flake.** (**a**) Tr-bb-TA spectrum of $T_d$-MoTe$_2$ recorded with $F$ = 0.5 mJ·cm$^{-2}$, $E_{pump}$ = 2.4 eV. Vibrational coherences are clearly observable as oscillatory components. (**b**) Black, temporal trace at $E_{probe}$ = 1.9 eV (horizontal line in panel **a**); red, linear fit performed within the interval $t$ = +2 – +14 ps to remove the electronic background and generate the residuals, $R_E$ (inset), for further FFT analysis. (**c**) Fourier power spectrum obtained from FFT of residuals showing the characteristic frequencies of the Raman active modes of the $T_d$ (1T'), $^1A_1$ ≈ 13 cm$^{-1}$, $^2A_1$ ($^1A_g$) ≈ 77 cm$^{-1}$ and $^5A_2$ ($^8A_g$) ≈ 164 cm$^{-1}$. (**d**) Fourier spectra as a function of $T$. These were obtained via averaging along $E_{probe}$ Fourier power spectra as the one shown in panel **c**.

Following the formation of $T^*$, we performed annealing (step 8 in Table 1) up to 500 K – the upper limit of our cryostat and found that $\mathcal{A}(^1A_1)$ does not recover (diamonds in Fig. 3b and step 8 in Table 1). This observation indicates that $T^*$ withstands thermal cycling, and therefore discards the possibility for $T^*$ to be a kinetically trapped $1T'$ state. The latter would have led, upon cooling, to the pristine $T_d$ structure with its characteristic $^1A_1$ vibrational coherence.

We then hypothesize that $T^*$ reflects a laser modified interlayer state, in which the coupling to the generation of $^1A_1$ vibrational coherences has been affected, for instance, through the formation of interlayer strain as recently observed in $T_d$-WTe$_2$ by FED[23]. To shine some light into the laser-induced formation of $T^*$, we performed laser irradiation tests with a limited number of shots at low repetition rate.

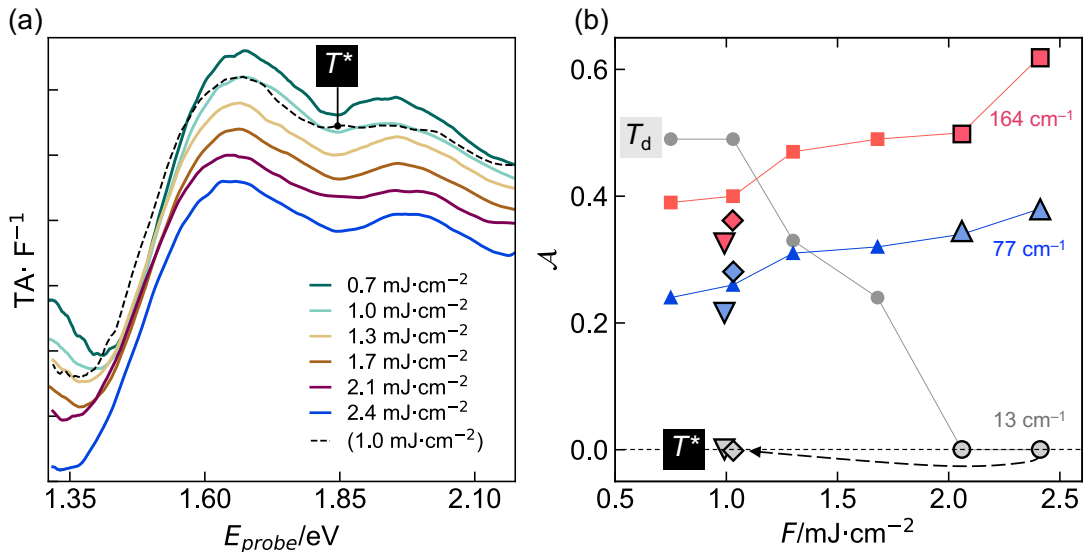

**Fig. 3: Fluence dependent measurements on a 32-nm thick $T_d$-MoTe$_2$ flake at a base temperature of 220 K.** (**a**) Fluence normalized transient absorption spectra (TA·$F^{-1}$) as a function of $F$. The parentheses indicate that $F$ was reduced to 1.0 mJ·cm$^{-2}$ after full transformation to $T^*$. The TA spectra were obtained by averaging the signal in the time domain between $t$ = +2 – +14 ps to wash out the effect of vibrational coherences. The spectra were offset for clarity. (**b**) $\mathcal{A}$ of three main coherent phonon modes. $\mathcal{A}(^1A_1)$ reaches background level at is at $F \approx 2$ mJ·cm$^{-2}$. At this point $T_d$ fully transformed into $T^*$. The diamonds correspond to a measurement under the same condition (i.e., 220 K) but after thermally annealing $T^*$ at 500 K and reducing $F$ to 1.0 mJ·cm$^{-2}$ as indicated by the arrow. The flake remained in the $T^*$ state. The inverted triangles correspond to a final measurement performed at 298 K and $F$ = 1.0 mJ·cm$^{-2}$.

**Table 1.** Measurements carried out on a 32-nm thick flake in chronological order from top to bottom (i.e., by step number). The higher number of scans at lower $F$ values were implemented to enhance the signal to noise ratio in our tr-bb-TA spectra.

| Specimen: 32-nm thick flake | | | | | | |
|---|---|---|---|---|---|---|
| Step | $F$ (mJ·cm$^{-2}$) | T (K) | scans | shots/scan | shots | Cumulative |
| 1 | 0.7 | 220 | 3 | 290,000,000 | 870,000,000 | 870,000,000 |
| 2 | 1.0 | 220 | 3 | 290,000,000 | 870,000,000 | 1,740,000,000 |
| 3 | 1.3 | 220 | 3 | 290,000,000 | 870,000,000 | 2,610,000,000 |
| 4 | 1.7 | 220 | 1 | 290,000,000 | 290,000,000 | 2,900,000,000 |
| 5 | 2.1 | 220 | 1 | 290,000,000 | 290,000,000 | 3,190,000,000 |
| 6 | 2.4 | 220 | 1 | 290,000,000 | 290,000,000 | 3,480,000,000 |
| 7 | Sample was annealed at 500 K and returned to 220 K for tr-bb-TA measurements. | | | | | |
| 8 | 1.0 (diamonds) | 220 | 3 | 290,000,000 | 870,000,000 | 4,350,000,000 |
| 9 | Sample was brought to room temperature for tr-bb-TA measurements. | | | | | |
| 10 | 1.0 (inverted triangles) | 298 | 3 | 290,000,000 | 870,000,000 | 5,220,000,000 |

The experiments shown in Fig. 4 are based on the exposure of thin flakes to a controlled number fs-laser shots at high $F$, followed by tr-bb-TA experiments carried out within the nondisruptive fluence regimen $F < 1$ mJ·cm$^{-2}$. Given the irreversible character of $T^*$, the latter measurements are used to monitor $\mathcal{A}(^1A_1)$; i.e., the presence or absence of the $^1A_1$ vibrational coherence. The values of $\mathcal{A}(^1A_1)$ have been normalized by the maximum magnitude of the TA signal to compensate for fluctuations in laser power arising from the sudden change of the repetition rate of our laser system between successive irradiation and test measurements as well as minor sample misalignments (see Supporting Information V).

Figure 4a exhibits an irradiation test performed on a 28-nm thick flake. The flake underwent several irradiation cycles while maintained at a base temperature of 200 K and at $F = 2.6$ mJ·cm$^{-2}$. This experiment approximately mimics the conditions for the 32-nm thick specimen in Fig. 3 at $F > 2$ mJ·cm$^{-2}$, but in the limit of fewer laser irradiation shots. Each black dot represents an irradiation cycle, and the number corresponds to the number of cumulative laser shots. The tr-bb-TA experiments necessary to determine $\mathcal{A}(^1A_1)$ were carry out at 100 K, i.e., the temperature of the sample was lowered and increased between irradiation cycles. The behavior of $\mathcal{A}(^1A_1)$ reveals that the formation process of $T^*$, i.e., strain buildup appears to be quite stochastic in the limit of a few to several thousands of shots. However, the fact that $\mathcal{A}(^1A_1)$ follows a decreasing trend indicates the gradual generation of $T^*$ with the increasing number of irradiation pulses.

Figure 4b shows the results obtained for a 35-nm thick flake exposed to a more extreme laser irradiation condition ($F$ = 6.5 mJ·cm$^{-2}$) while maintained at a base temperature of 85 K. Each of the first five irradiation cycles involved 10 laser shots. $\mathcal{A}(^1A_1)$ was found to experience a decrease of ≈ 25% from the first to the second irradiation round, remained nearly constant during the next three irradiation cycles, and suddenly approached the background level after the fourth exposure. We monitored the maximum magnitude of the TA signal as well as $\mathcal{A}(^2A_1)$ during our tr-bb-TA experiments to discard the possibility of excessive sample damage before continuing with the next irradiation round. Note that $F$ = 6.5 mJ·cm$^{-2}$ is within the $F_{th}$ range for film damage, *vide supra*. We did not observe signs of sample degradation as evidenced in other laser irradiation tests (Supporting Information V, test 3). The specimen was then warmed up to 298 K, exposed to 100 additional fs-laser shots with $F$ = 4.6 mJ·cm$^{-2}$, and brought back to 85 K to perform tr-bb-TA measurements. The clear recovery of $\mathcal{A}(^1A_1)$ not only confirms a decent level of sample integrity but also evidences that impulsive fs-laser irradiation at room temperature can be employed to transform $T^*$ into the pristine-like $1T'$ phase which, upon cooling, transitions to the pristine $T_d$ state.

Furthermore, we also discovered that laser irradiation of the high temperature $1T'$ state does not lead to the formation of $T^*$, see Fig. S11. On the contrary, impulsive laser irradiation of the $1T'$ was found, in some cases, to slightly enhance $\mathcal{A}(^1A_1)$ when the sample is cooled down to the $T_d$ state, which is aligned with our hypothesis of laser induced interlayer strain formation. We think that the latter observation arises from the fact that pristine flakes carry some residual level of strain introduced during the transfer and h-BN capping processes of the $1T'$ MoTe$_2$ film. The formation of a strained state is also justified by *ex-situ* transport measurements we performed as a function of temperature before and after laser irradiation. Measurements show that $T^*$ undergoes hysteresis upon thermal cycling and it is comparable to that observed for the pristine state (see Supporting Information VII). This observation suggests that the structures of $T^*$ at low and high temperatures may correlate with the strained structures of $T_d^*$ and $1T'^*$ phases, respectively. However, this requires further investigation.

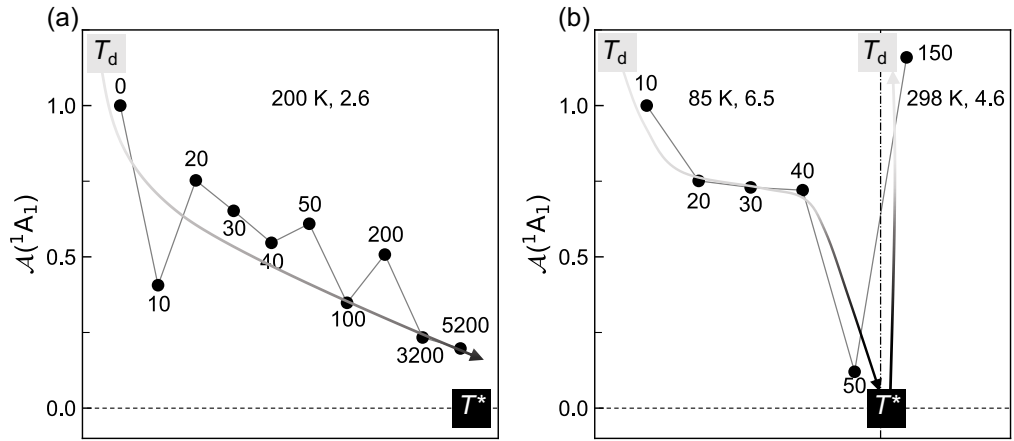

Fig. 4: Formation of $T^*$ under few-shots photoexcitation conditions and its reversion at room temperature to $1T'$ (i.e., $T_d$ upon cooling). (a, b) $\mathcal{A}(^1A_1)$ as a function of the cumulative number of laser shots from left to right as indicated next to each data point. $\mathcal{A}(^1A_1)$ has been normalized by the maximum magnitude of the TA signal, which served as an internal power reference (see Fig. S6). (a) 28-nm thick flake irradiated at a base temperature of 200 K and $F \approx 2.6$ mJ·cm$^{-2}$ (units of $F$ where not included in panels). The sample was brought to 100 K after each irradiation cycle to run tr-bb-TA (b) Similar to panel **a** but for a 35-nm thick flake. For the first 50 shots the sample was kept at a base temperature of 85 K to perform each irradiation cycle at $F \approx 6.5$ mJ·cm$^{-2}$. Vibrational coherences were also measured at 85 K. The vertical dashed line indicates that the sample was brought to 298 K, irradiated by additional 100 shots at $F \approx 4.6$ mJ·cm$^{-2}$ and cooled down to 85 K to measure again vibrational coherences at low $F$. Tr-bb-TA experiments were carried out at $F \approx 0.5$ mJ·cm$^{-2}$. The pump photon energy was $E_{pump} = 2.4$ eV for both irradiation and tr-bb-TA measurements. The arrows are guides to eye in the formation of and reversion from $T^*$.

It is interesting to speculate about the mechanism driving the formation of $T^*$. The fact that $T^*$ was found to only develop upon fs-laser irradiation of the $T_d$ state suggests that the process of interlayer strain formation could be due to a frustrated photoinduced $T_d \rightarrow 1T'$ phase transformation. This interpretation is in agreement with the small interlayer displacement unveiled in the same compound by FED[19] and the recent observation of a transiently hot metastable $T_d$ phase via bb-spectroscopy[20].

Moreover, the fact that such persistent $T^*$ phase is observed in semi-freestanding thin flakes and not in a bulk crystal[20] suggests that the difference in the cooling rate of such metastable hot $T_d$ state[20] may play a role as well as the interactions of the thin flake with the thin $Si_3N_4$ substrate and BN capping layer. Note that this effect was not observed in the freestanding $T_d$-$MoTe_2$ flakes during FED experiments[19]. Each fs-laser pulse deposits energy and leads to the transient photoinduced increase of the lattice temperature well above $T_c$ (see Supporting Information III). This impulsive heating effect is followed by cooling of the sample, which should reach the base temperature imposed by the cryostat before the arrival of the next fs-pump pulse. The main difference between a semi-freestanding thin flake and a bulk crystal comes from their cooling efficiency, which is about $10^3$ times faster for the latter. In the case of our flakes, their semi-freestanding geometry substantially reduces the overall rate for thermal energy transfer to the substrate. Note that flakes would quickly but only partially thermalize with the thin BN and $Si_3N_4$ layers. However, the final heat transfer to the thick Si frame occurs along the plane of the layers. Such in-plane or transverse heat transfer, as opposed to the longitudinal energy transfer experienced by the hot surface of a bulk crystal, is known to proceed on the timescale of several hundreds of microseconds; see Supplementary Information in reference 4. We believe that this extended cooling period is necessary for the transiently hot $T_d$ lattice to be able develop into the observed strained $T^*$ state.

The formation of light-induced long-live states of matter and ultrafast PIPTs[37,38] are playing an increasingly important role in quantum materials and devices[3,39] designed to confer properties on demand[40]. Our study reveals that impulsive laser irradiation can be utilized to induce or release interlayer strain that is likely arising from a frustrated (incomplete) photoinduced phase transformation. It provides the opportunity to extend such phenomenon to other 2D-TMDCs such as $T_d$-WTe$_2$ for which THz-light induced strained formation has recently shown to create a topologically distinct metastable phase[23]. Follow-up experiments will involve the implementation of THz-Raman and FED with transport measurement capabilities *in situ* to bring new vistas in the understanding of the structural and electronic changes induced by intense fs-laser irradiation in a growing body of novel 2D materials.

**Acknowledgements**

We would like to thank Professor Roberto Merlin (U. Michigan) and Dr. Ralph Ernstorfer (Fritz Haber Institute) for fruitful discussions and Professor David Cory (U. Waterloo) for lending us the Oxford cryostat utilized in this work. A.W.T. acknowledges support from the U.S. Army Research Office (No. W911NF-21-2-0136). A.W.T. and G.S. acknowledge the support of the National Science and Engineering Research Council of Canada and the Ontario Early Researcher Award program. GS also acknowledges the support of the Canada Foundation for Innovation, and Canada Research Chair program. This research was undertaken thanks in part to funding from the Canada First Research Excellence Fund. F.C., X.L., and Y.S. thank the support from the National Key Research and Development Program under Contract No. 2021YFA1600201, the National Nature

**Competing interests**

**Data availability**

The data that support the findings of this study are available from the corresponding author upon reasonable request.

**Correspondence and requests for materials** should be addressed to G.S. (gsciaini@uwaterloo.ca).

**References**

1. Strickland, D. & Mourou, G. Compression of amplified chirped optical pulses. *Opt. Commun.* **56**, 219–221 (1985).
2. Nasu, K. *Photoinduced Phase Transitions*. (World Scientific, 2004).
3. Stojchevska, L. *et al.* Ultrafast Switching to a Stable Hidden Quantum State in an Electronic Crystal. *Science* **344**, 177–180 (2014).
4. Eichberger, M. *et al.* Snapshots of cooperative atomic motions in the optical suppression of charge density waves. *Nature* **468**, 799–802 (2010).
5. Han, T.-R. T. *et al.* Exploration of metastability and hidden phases in correlated electron crystals visualized by femtosecond optical doping and electron crystallography. *Sci. Adv.* **1**, e1400173 (2015).
6. Vogelgesang, S. *et al.* Phase ordering of charge density waves traced by ultrafast low-energy electron diffraction. *Nat. Phys.* **14**, 184–190 (2018).
7. Vellinga, M. B., de Jonge, R. & Haas, C. Semiconductor to metal transition in $MoTe_2$. *J. Solid State Chem.* **2**, 299–302 (1970).
8. Rhodes, D. *et al.* Engineering the Structural and Electronic Phases of $MoTe_2$ through W Substitution. *Nano Lett.* **17**, 1616–1622 (2017).
9. Song, S. *et al.* Room Temperature Semiconductor–Metal Transition of $MoTe_2$ Thin Films Engineered by Strain. *Nano Lett.* **16**, 188–193 (2016).
10. Wang, L. *et al.* One-Dimensional Electrical Contact to a Two-Dimensional Material. *Science* **342**, 614–617 (2013).

11. He, R. *et al.* Dimensionality-driven orthorhombic MoTe$_2$ at room temperature. *Phys. Rev. B* **97**, 041410 (2018).

12. Soluyanov, A. A. *et al.* Type-II Weyl semimetals. *Nature* **527**, 495–498 (2015).

13. Jiang, J. *et al.* Signature of type-II Weyl semimetal phase in MoTe$_2$. *Nat. Commun.* **8**, 13973 (2017).

14. Yan, B. & Felser, C. Topological Materials: Weyl Semimetals. *Annu. Rev. Condens. Matter Phys.* **8**, 337–354 (2017).

15. Zhang, K. *et al.* Raman signatures of inversion symmetry breaking and structural phase transition in type-II Weyl semimetal MoTe$_2$. *Nat. Commun.* **7**, 13552 (2016).

16. Chen, S.-Y., Goldstein, T., Venkataraman, D., Ramasubramaniam, A. & Yan, J. Activation of New Raman Modes by Inversion Symmetry Breaking in Type II Weyl Semimetal Candidate T'-MoTe$_2$. *Nano Lett.* **16**, 5852–5860 (2016).

17. Zhang, M. Y. *et al.* Light induced sub-picosecond topological phase transition in MoTe$_2$. *ArXiv180609075 Cond-Mat* (2018).

18. Zhang, M. Y. *et al.* Light-Induced Subpicosecond Lattice Symmetry Switch in MoTe$_2$. *Phys. Rev. X* **9**, 021036 (2019).

19. Qi, Y. *et al.* Photoinduced concurrent intralayer and interlayer structural transitions and associated topological transitions in MTe$_2$ (M=Mo, W). *ArXiv210514175 Cond-Mat* (2021).

20. Cheng, M. *et al.* Photoinduced interlayer dynamics in T$_d$-MoTe$_2$: A broadband pump-probe study. *Appl. Phys. Lett.* **120**, 123102 (2022).

21. Sciaini, G. & Miller, R. J. D. Femtosecond electron diffraction: heralding the era of atomically resolved dynamics. *Rep. Prog. Phys.* **74**, 096101 (2011).

22. Sciaini, G. Recent Advances in Ultrafast Structural Techniques. *Appl. Sci.* **9**, 1427 (2019).

23. Sie, E. J. *et al.* An ultrafast symmetry switch in a Weyl semimetal. *Nature* **565**, 61 (2019).

24. Mannebach, E. M. *et al.* Dynamic Optical Tuning of Interlayer Interactions in the Transition Metal Dichalcogenides. *Nano Lett.* **17**, 7761–7766 (2017).

25. Kumazoe, H. *et al.* Photo-induced lattice contraction in layered materials. *J. Phys. Condens. Matter* **30**, 32LT02 (2018).

26. Cavalleri, A. *et al.* Femtosecond Structural Dynamics in VO$_2$ during an Ultrafast Solid-Solid Phase Transition. *Phys. Rev. Lett.* **87**, 237401 (2001).

27. Cavalleri, A., Dekorsy, Th., Chong, H. H. W., Kieffer, J. C. & Schoenlein, R. W. Evidence for a structurally-driven insulator-to-metal transition in VO$_2$: A view from the ultrafast timescale. *Phys. Rev. B* **70**, 161102 (2004).

28. Baum, P., Yang, D.-S. & Zewail, A. H. 4D Visualization of Transitional Structures in Phase Transformations by Electron Diffraction. *Science* **318**, 788–792 (2007).

29. Tao, Z. *et al.* Decoupling of Structural and Electronic Phase Transitions in VO$_2$. *Phys. Rev. Lett.* **109**, 166406 (2012).

30. Wall, S. *et al.* Ultrafast changes in lattice symmetry probed by coherent phonons. *Nat. Commun.* **3**, 721 (2012).

31. Wall, S. *et al.* Ultrafast disordering of vanadium dimers in photoexcited $VO_2$. *Science* **362**, 572–576 (2018).

32. Morrison, V. R. *et al.* A photoinduced metal-like phase of monoclinic $VO_2$ revealed by ultrafast electron diffraction. *Science* **346**, 445–448 (2014).

33. Gao, M. *et al.* Mapping molecular motions leading to charge delocalization with ultrabright electrons. *Nature* **496**, 343–346 (2013).

34. Ishikawa, T. *et al.* Direct observation of collective modes coupled to molecular orbital–driven charge transfer. *Science* **350**, 1501–1505 (2015).

35. Rivas, N. *et al.* Generation and detection of coherent longitudinal acoustic waves in ultrathin 1T'-$MoTe_2$. *Appl. Phys. Lett.* **115**, 223103 (2019).

36. Santos-Cottin, D. *et al.* Low-energy excitations in type-II Weyl semimetal $T_d$-$MoTe_2$ evidenced through optical conductivity. *Phys. Rev. Mater.* **4**, 021201 (2020).

37. Zhang, J. *et al.* Cooperative photoinduced metastable phase control in strained manganite films. *Nat. Mater.* **15**, 956–960 (2016).

38. Li, X. *et al.* Terahertz field–induced ferroelectricity in quantum paraelectric $SrTiO_3$. *Science* **364**, 1079–1082 (2019).

39. Cho, S. *et al.* Phase patterning for ohmic homojunction contact in $MoTe_2$. *Science* **349**, 625–628 (2015).

40. Basov, D. N., Averitt, R. D. & Hsieh, D. Towards properties on demand in quantum materials. *Nat. Mater.* **16**, 1077–1088 (2017).

# Supporting Information

# Persistent photogenerated state attained by femtosecond laser irradiation of thin $T_d$-MoTe$_2$

Meixin Cheng[1], Shazhou Zhong[2], Nicolas Rivas[1,♦], Tina Dekker[2], Ariel Alcides Petruk[1], Patrick Gicala[1], Kostyantyn Pichugin[1], Fangchu Chen[2], Xuan Luo[3], Yuping Sun[3,4,5], Adam W. Tsen[2], Germán Sciaini[1,*]

[*]correspondence: gsciaini@uwaterloo.ca

[♦]Present address: Nuclear Engineering Group, McMaster University, Hamilton, Ontario L8S 4K1, Canada.

[*]Correspondence: gsciaini@uwaterloo.ca

[1]The Ultrafast Electron Imaging Lab, Department of Chemistry, and Waterloo Institute for Nanotechnology, University of Waterloo, Waterloo, Ontario N2L 3G1, Canada.

[2]Institute for Quantum Computing, Department of Physics and Astronomy, Department of Electrical and Computer Engineering, and Department of Chemistry, University of Waterloo, Waterloo, Ontario N2L 3G1, Canada.

[3]Key Laboratory of Materials Physics, Institute of Solid-State Physics, HFIPS, Chinese Academy of Sciences, Hefei 230031, China.

[4]High Magnetic Field Laboratory, Chinese Academy of Sciences, HFIPS, Hefei 230031, China.

[5]Collaborative Innovation Center of Advanced Microstructures, Nanjing University, Nanjing 210093, China.

## I. Experimental Methods

**Broadband fs-transient absorption/reflection measurements.** Our laser system is from Light Conversion, model Pharos SP and delivers 300 µJ, 170 fs pulses centered at the fundamental wavelength of 1030 nm (1.2 eV). The maximum repetition rate of our system is 20 kHz and tr-bb-TA experiments were performed, depending on flake thickness, between 1 kHz and 20 kHz. An optical parametric amplifier followed by a second-harmonic generation module yields pulses of > 5 µJ, 100 fs, centred at 520 nm (2.4 eV). We utilized pump pulses with $E_{pump}$ = 2.4 eV and ~100-fs (full-width-at-half-maximum, fwhm) in duration to perform all tr-bb-TA in flakes. This $E_{pump}$ was outside the $E_{probe}$ window and facilitated the alignment of the pump beam by monitoring the transmission of green light through the body of each flake deposited onto each small 10-um x 10-um $Si_3N_4$ window. Broadband probe pulses were generated by focusing a small fraction of the fundamental beam into a 3-mm thick YAG crystal using a 50-mm focal length lens. Subsequently, a short pass filter was implemented to remove the residual fundamental avoiding saturation on the spectrometer and leaving a usable $E_{probe}$ 1.3 eV to 2.1 eV. The white-light beam was collimated by an off-axis parabolic mirror and afterwards focused using a 50 mm focal length lens. The probe beam spot size at the sample was about 20 µm (fwhm) whereas the pump beam spot size was set to 350 µm (fwhm). The TA experiments were performed in a quasi-collinear arrangement with an angle of 10° between the incident pump and probe beams. The incident pump fluence was varied through the combination of a half-wave plate and a polarizer. Differential bb-TA spectra were obtained by modulating the pump beam with a mechanical chopper at 42 Hz. An optical delay stage was used to control the time delay between pump and probe pulses. The tr-bb-TA spectra were recorded using a synchronized dispersive spectrometer. The sample was supported in an 8 mm x 8 mm silicon frame which fits in the vacuum sealed compartment of an optical cryostat, which controls the temperature from 75 K up to 500 K. Once the system reached stable conditions, temperature fluctuations were within ± 0.01 K. Data analyses were performed in Python 2.7.

**Sample Preparation.** Ultrathin flakes were obtained by mechanical exfoliation of a $1T''$-MoTe$_2$ single crystal inside a nitrogen-filled glove box and transferred onto specially nanofabricated substrates with a 10-um x 10-um 50-nm thick Si$_3$N$_4$ window. MoTe$_2$ films prepared by mechanical exfoliation of a single crystal are known to be of high crystalline quality and have been previously shown to retain their metallic properties down to low temperatures[11,43]. The freestanding Si$_3$N$_4$ windows served to spatially overlap the beams and to ensure that the entire transmission of the probe beam passed through the MoTe$_2$ film of interest. It was essential to assure the whole window was covered by the flake under study. Hexagonal BN was used to cap the $1T''$-MoTe$_2$ for protection against ambient moisture. The main challenge in our tr-bb-TA studies has been to achieve high signal to noise ratio (SNR) in transient spectra attained from such small, probed volumes, and for thicker flakes owing to the poorer probe beam transmittance of samples that are semimetallic in nature. This limited the total timespan in our tr-bb-TA experiments to about 15 ps. This experimental configuration, however, guaranteed the characterization of a well-defined body; reproducible positioning of the flake and enabled the performance of sequential studies, e.g. tr-bb-TA measurements as a function of base temperature, after thermal annealing, irradiation cycles, etc. on a given specimen. Note that owing to thermal expansion effects, a change of the base temperature led to sample drift, and therefore the proper repositioning of the specimen becomes crucial to succeed with comparative analyses. In addition, experiments should not last for more than a week since MoTe$_2$ is sensitive to moisture and degrades.

## II. Data Fitting Procedure

**Chirp correction**

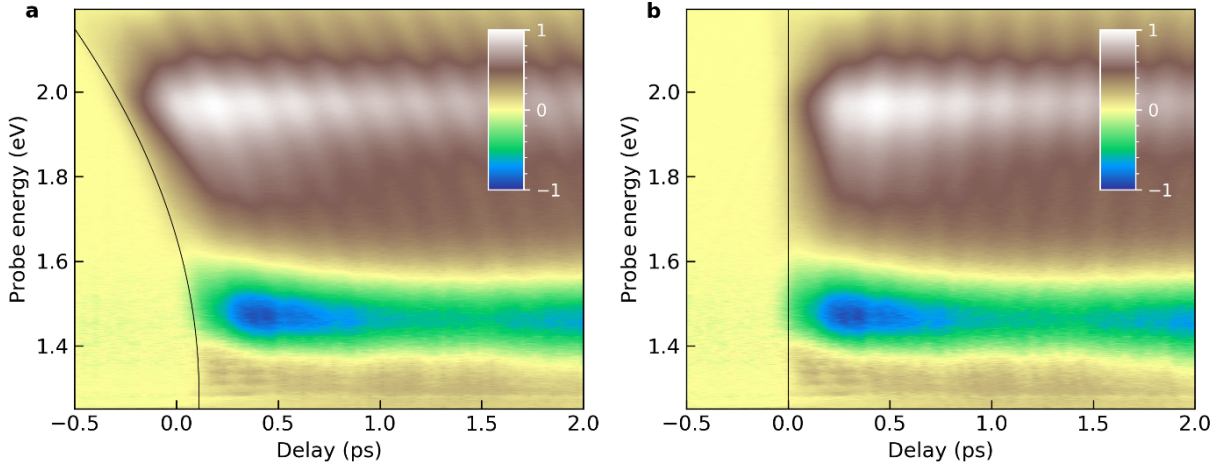

**Fig. S1: Chirp correction procedure. (a)** Raw tr-bb-TA spectra of 40-nm thick $T_d$-MoTe$_2$ flake at 74 K. **(b)** Chirp-corrected tr-bb-TA spectra. The broadband probe pulses carry positive chirp since they transverse some dispersive materials (e.g. lenses, cryostat window, etc.) before reaching the specimen. The chirp was corrected by fitting a polynomial to the TA-signal onset and time-shifting each probe energy cross section to match the same origin or time zero. The thin solid black lines represent the time-zero curve before **a** and after **b** chirp correction.

**Automated fitting procedure for the generation of residuals**

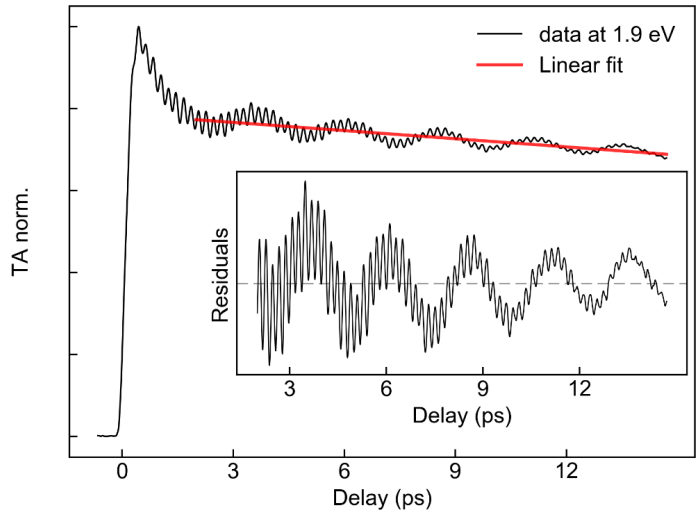

**Fig. S2: Example of an automated fitting procedure.** A time trace (black) at $E_{probe}$ = 1.9 eV is fitted via a linear regression within the time delay interval $t$ = +2 ps – +14 ps. The result from the linear fit is in red. Sample, 40-nm thick $T_d$-MoTe$_2$; temperature = 74 K. The residual trance is shown in the inset.

Different fitting procedures were carried out to remove the dynamic electronic background from the tr-bb-TA spectra and generate residuals for the analysis of coherent phonon dynamics. To avoid arbitrariness and enable the automation and the analyses of large data sets, we decided to fit temporal traces by an exponential or a linear trend for time delays, $t > 2$ ps. For $t > 2$ ps electronic driven changes of the TA signal have reached a quasi-steady state and time-dependent modulations are mostly governed by coherent phonon oscillations. Both fitting functions delivered very similar results (see Fig. 2b in manuscript).

### III. Photoinduced Temperature Calculation

Upon laser excitation the thin flake suffers an increment in its lattice temperature, which can be estimated via a simple energy balance, according to Eq. S1:

$$\int_{T_0}^{T_f} C_P(T)\, dT = \frac{F\, A\, M}{\rho\, L} \qquad \text{(Eq. S1)}$$

Where $C_p$ is the heat capacity of MoTe$_2$[1], $T$ is the temperature, $T_0$ is the initial or base temperature defined by the cryostat, $T_f$ is the final photoinduced temperature, $F$ is the incident fluence, $M$ is the molar mass of MoTe$_2$ (351.1 g·mol$^{-1}$), $\rho$ is the density (7.78 g·cm$^{-3}$), $L$ is the thickness of the MoTe$_2$ flake, which is usually comparable to the optical absorption depth of the pump, $\delta \approx 32$ nm at $E_{\text{pump}} = 2.4$ eV. $A$ is the absorbed fraction of incident laser power by the MoTe$_2$ flake that is estimated via Eq. S2 using Fresnel equations to account for the reflections in cryostat window, the BN layer and the BN-MoTe$_2$ interface, providing a value of $R = 0.28$.

$$A = 1 - R - (1 - R)\, exp(-L/\delta) \qquad \text{(Eq. S2)}$$

Thus, the absorbed fraction for a $T_d$-MoTe$_2$ flake with a thickness of 32 nm results to be $A = 0.43$.

We also estimated the amount of absorbed fluence, $F_{\Delta H}$, necessary to overcome the latent heat of reaction, $\Delta H_{Td-1T'}$. Since we lack a value for the $T_d$–$1T'$ phase transformation, we used the value that was determined for the 2H–$1T'$ phase transition[3] $\Delta H_{2H\text{-}1T'} = 360$ cal·mol$^{-1}$. Note that the $T_d$–$1T'$ phase transition involves a smaller structural change, and therefore it is expected that $\Delta H_{Td-1T'} < \Delta H_{2H\text{-}1T'}$ and thus,

$$F_{\Delta H_{2H-1T'}} = \frac{\Delta H_{2H-1T'}\,\rho\, L}{M} \qquad (\text{Eq. S4})$$

could be seen as an upper estimate of $F_{\Delta H_{Td-1T'}}$, which for the 32-nm flake results to be only $F_{\Delta H_{2H-1T'}} \approx 0.11$ mJ·cm$^{-2}$. Note that the value of $F_{\Delta H_{2H-1T'}}$ is a small fraction of the employed $F$ or absorbed fluence, and therefore it can be neglected.

**Table S1.** Estimated maximum photoinduced temperature as function of $F$ at two different initial temperatures for a generic 30-nm thick $T_d$-MoTe$_2$ flake.

| 30-nm thick flake | | | | |
|---|---|---|---|---|
| $F$ | $T_0$ | $T_{f,\text{photoinduced}}(K)$ | $T_0$ | $T_{f,\text{photoinduced}}(K)$ |
| 1.0 | 85 | 189 | 220 | 307 |
| 2.0 | 85 | 272 | 220 | 380 |
| 3.0 | 85 | 338 | 220 | 450 |
| 4.0 | 85 | 385 | 220 | 517 |
| 5.0 | 85 | 414 | 220 | 581 |

## IV. Fourier Power Spectra for Flakes under Different Conditions

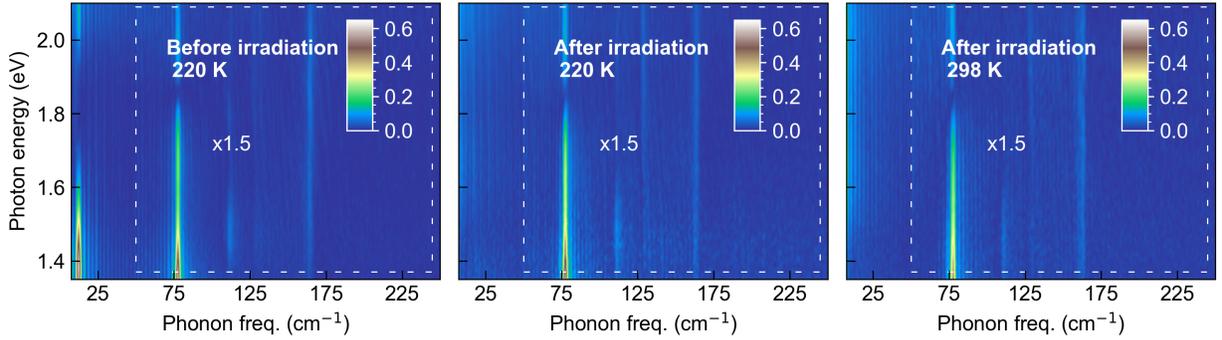

**Fig. S3: Fourier power spectra obtained for a 32-nm thick MoTe$_2$ film before and after intense irradiation at 20 kHz.** Left panel, pristine sample below $T_c$. Central panel, sample after irradiation with $F$ = 2.4 mJ·cm$^{-2}$ below $T_c$. The $^1A_1$ coherence is no longer present below $T_c$. Right panel, sample after irradiation with $F$ = 2.4 mJ·cm$^{-2}$ above $T_c$ (presumably 1$T$'). Power spectra recorded at $F$ = 1.0 mJ·cm$^{-2}$ and $E_{pump}$ = 2.4 eV. Tr-bb-TA measurements were performed at 20 kHz.

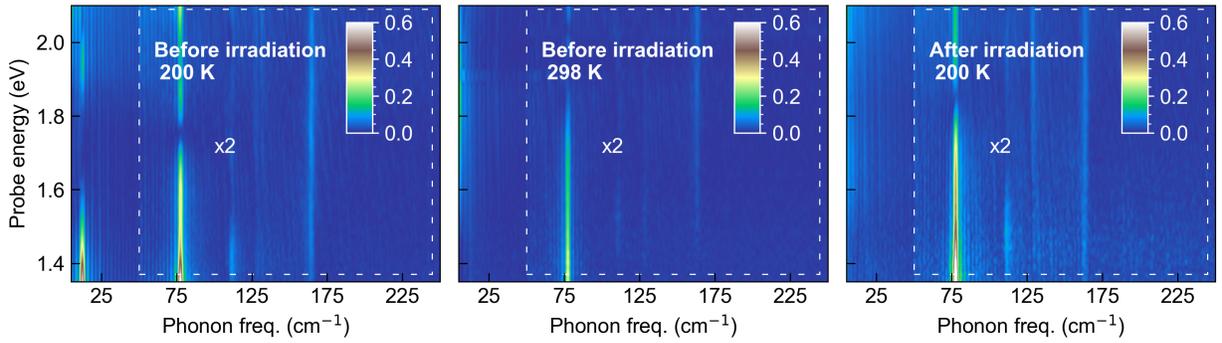

**Fig. S4: Fourier power spectra obtained for a 30-nm thick MoTe$_2$ film before and after irradiation.** Left panel, pristine $T_d$-MoTe$_2$ sample at 200 K. Central panel, pristine 1$T$' sample at 298 K. Right panel, after 5 minutes of laser irradiation at 10 Hz, 85 K, $F$ = 4.2 mJ·cm$^{-2}$. The coherent $^1A_1$ phonon mode ≈ 13 cm$^{-1}$ vanishes. 2D power spectra were recorded at $F$ = 0.6 mJ·cm$^{-2}$ and $E_{pump}$ = 2.4 eV. Tr-bb-TA measurements were performed at 20 kHz.

## V. Laser Irradiation Tests in Thin Flakes

Laser irradiation tests were performed on three flakes to study the effects of laser-induced interlayer strain buildup and sample damage under controlled laser irradiation conditions. The experiments shown below are based on the exposure of thin flakes to controlled fs-laser irradiation at high $F$ and low repetition rate (10 Hz) followed by tr-bb-TA measurements carried out within the nondisruptive $F$ regimen < 1 mJ·cm$^{-2}$ at 20 kHz. Given the persistent character of $T^*$, our tr-bb-TA measurements served to assess the state of film by monitoring the $^1A_1$ vibrational coherence.

**Test 1: 35-nm thick MoTe$_2$ sample – strain buildup and release**

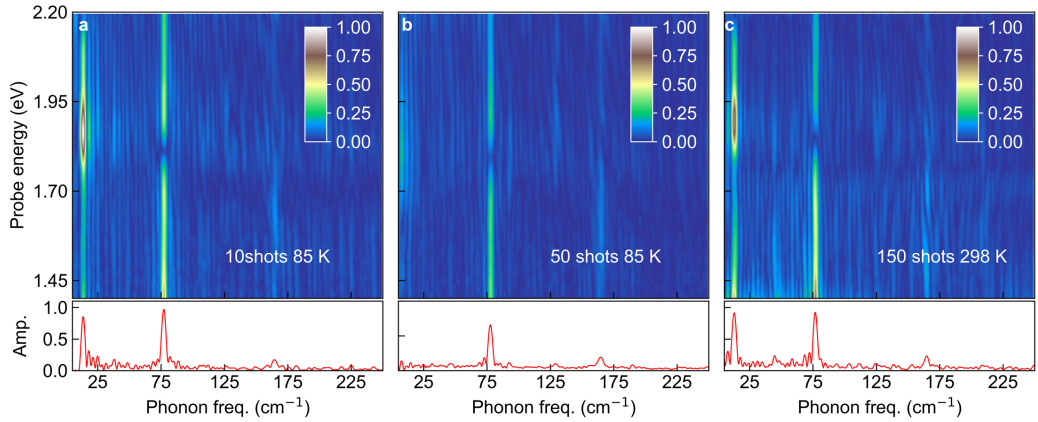

**Fig. S5: Fourier power spectra.** (**a**) After laser irradiation, 10 shots at 85 K, $F$ = 6.5 mJ·cm$^{-2}$. (**b**) Same specimen after a total of 50 shots at 85 K, $F$ = 6.5 mJ·cm$^{-2}$. (**c**) Same specimen after additional 100 shots at 298 K, $F$ = 4.6 mJ·cm$^{-2}$. All tr-bb-TA were performed at 85 K, $F$ = 0.5 mJ·cm$^{-2}$. All data were normalized by the maximum FFT amplitude obtained for the experiment shown in panel **a**.

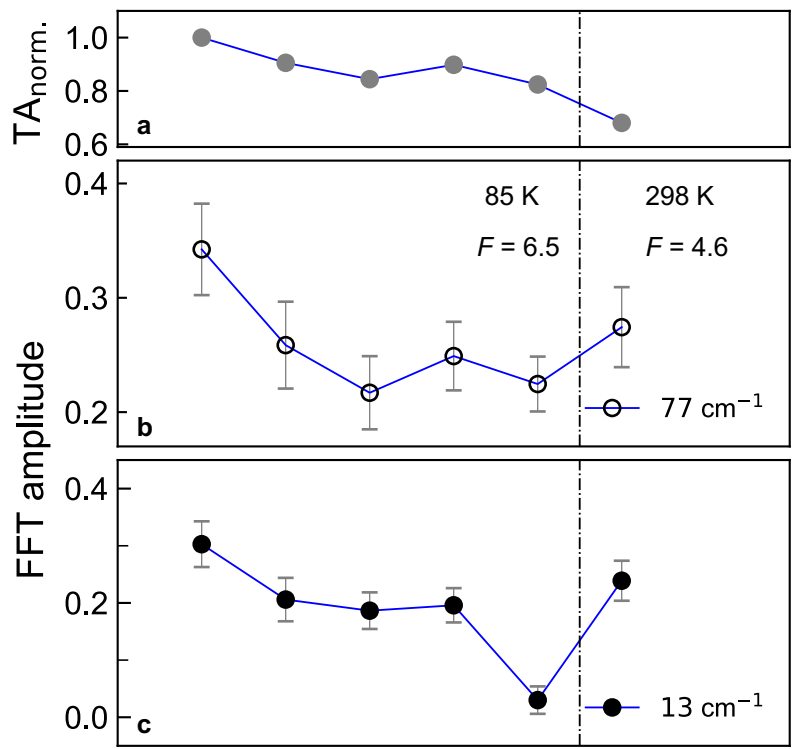

**Fig. S6: Laser induced changes as a function of the cumulative number of laser irradiation pulses.** From left to right the number of cumulative laser shots is 10, 20, 30, 40, 50 and 150. **(a)** Recorded maximum magnitude of TA signal, normalized by its initial value. **(b)** FFT-amplitude of the 77 cm$^{-1}$ mode. **(c)** FFT-amplitude of the 13 cm$^{-1}$ mode. The temperatures and fluences employed during laser irradiation exposures are indicated in the figure. All tr-bb-TA measurements were carried out at 85 K, $F \approx 0.5$ mJ·cm$^{-2}$.

**Test 2: 28-nm thick MoTe$_2$ sample – strain buildup**

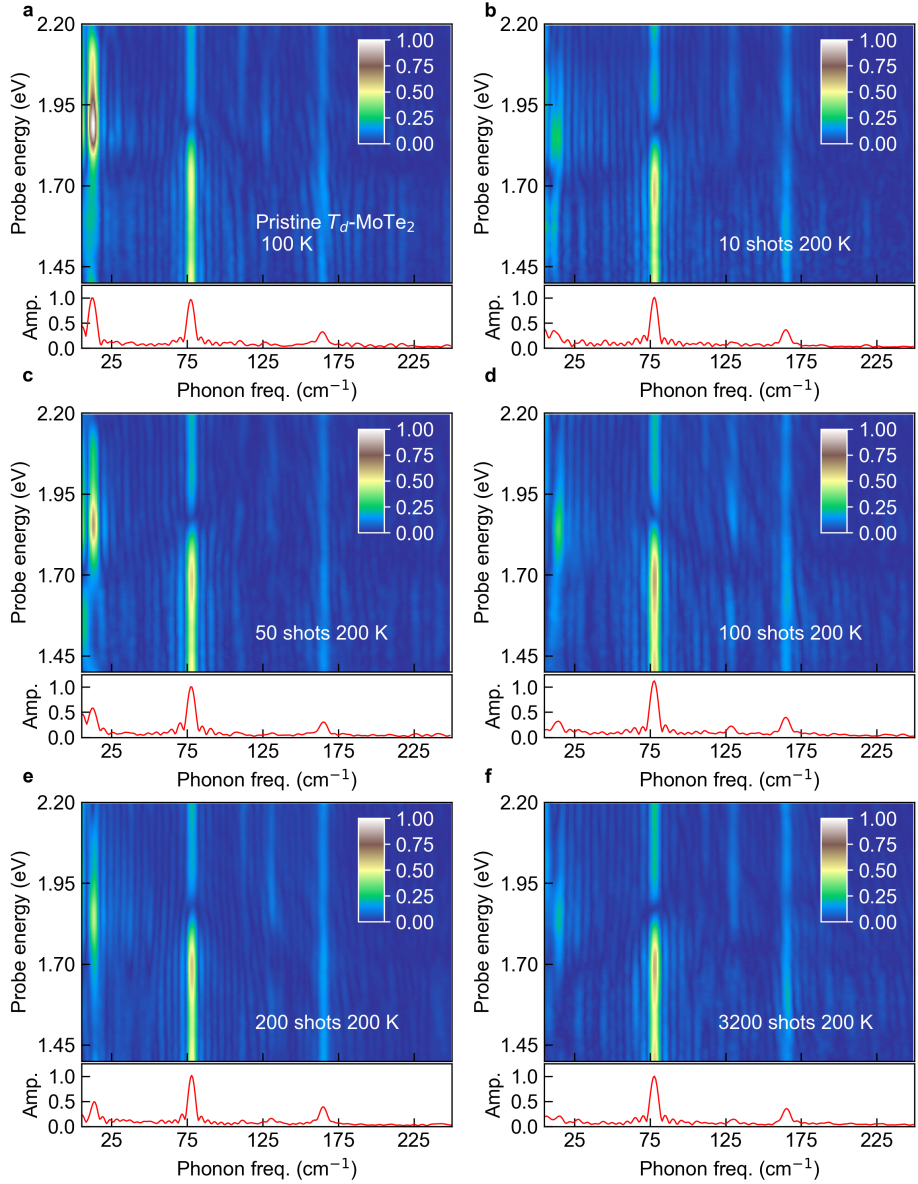

**Fig. S7: Fourier power spectra. (a)** Pristine sample. **(b-f)** Same specimen following laser irradiation exposures. The temperatures and cumulative number of laser shots employed in irradiation cycles are indicated in each panel. Irradiation exposures were carried out with $F = 2.6$ mJ·cm$^{-2}$. All tr-bb-TA measurements performed at 100 K, $F \approx 0.4$ mJ·cm$^{-2}$.

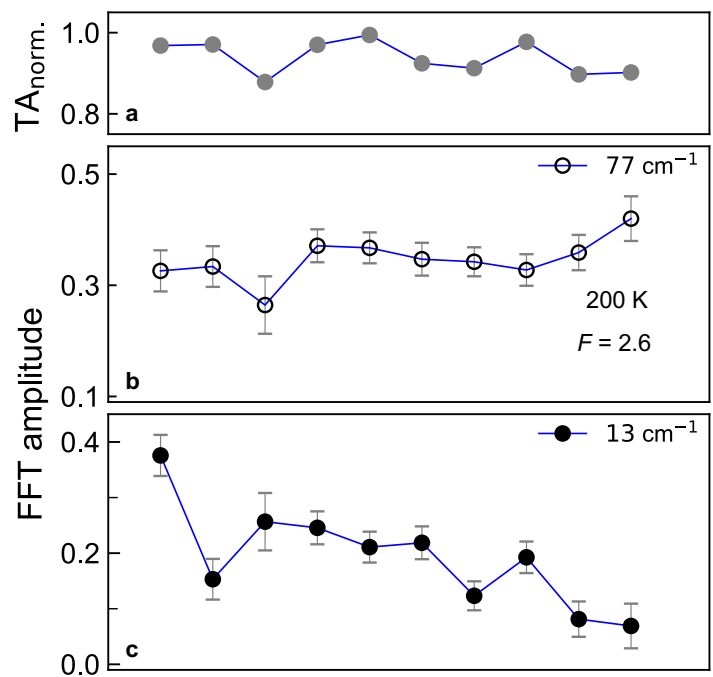

**Fig. S8: Laser induced changes as a function of the cumulative number of laser irradiation pulses.** From left to right the number of cumulative laser shots is 0, 10, 20, 30, 40, 50, 100, 200, 3200, and 5200. **(a)** Recorded maximum magnitude of TA signal, normalized by its initial value. **(b)** FFT-amplitude of the 77 cm$^{-1}$ mode. **(c)** FFT-amplitude of the 13 cm$^{-1}$ mode. The temperatures and fluences employed during laser irradiation exposures are indicated in the figure. All tr-bb-TA measurements were carried out at 100 K, $F \approx 0.4$ mJ·cm$^{-2}$.

**Test 3: 35-nm thick MoTe$_2$ sample – film damage**

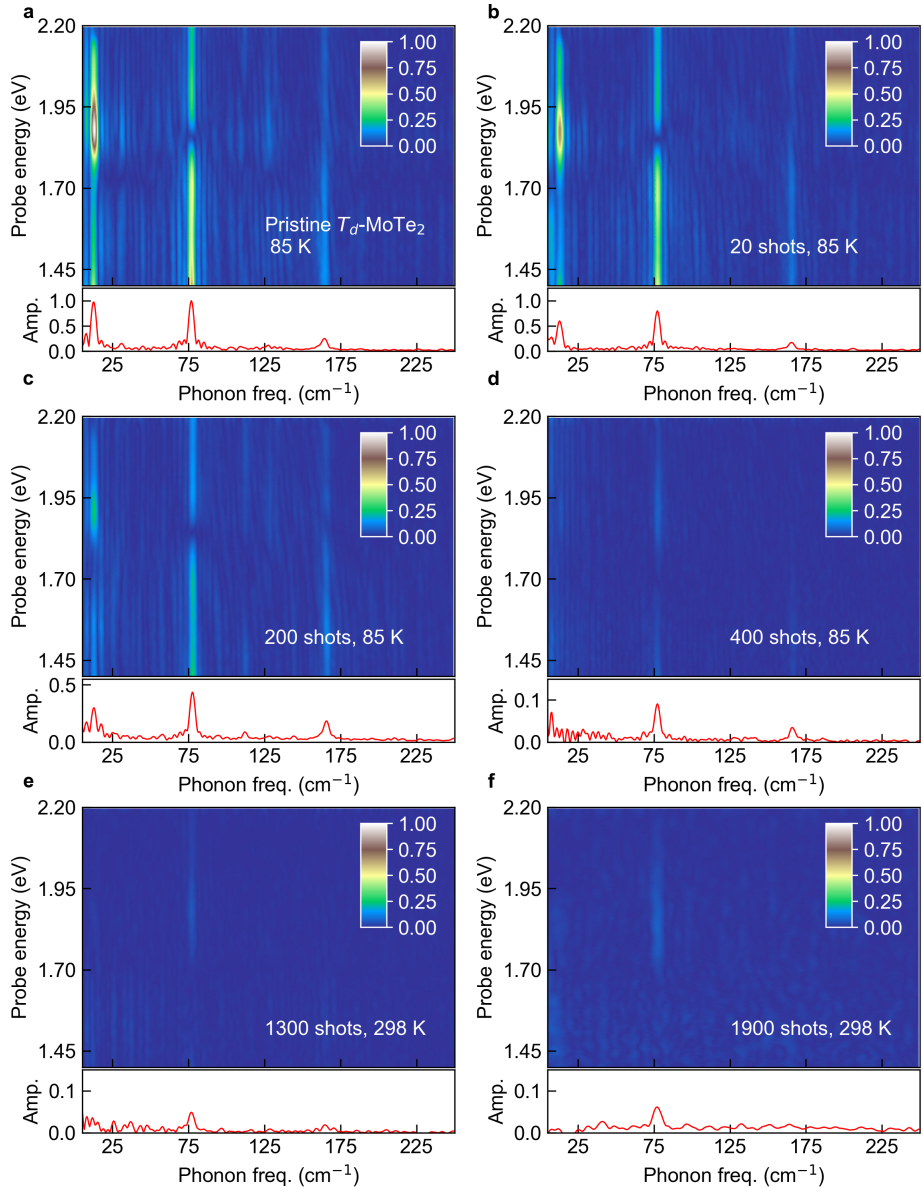

**Fig. S9: Fourier phonon power spectra. (a)** Pristine sample. **(b-f)** Same specimen following laser irradiation exposures. The temperatures and cumulative number of laser shots employed in irradiation cycles are indicated in each panel. Irradiation exposures were carried out with $F$ = 4.6 mJ·cm$^{-2}$. All tr-bb-TA measurements performed at 85 K, $F \approx 0.4$ mJ·cm$^{-2}$. There is a clear sign of sample degradation (see below).

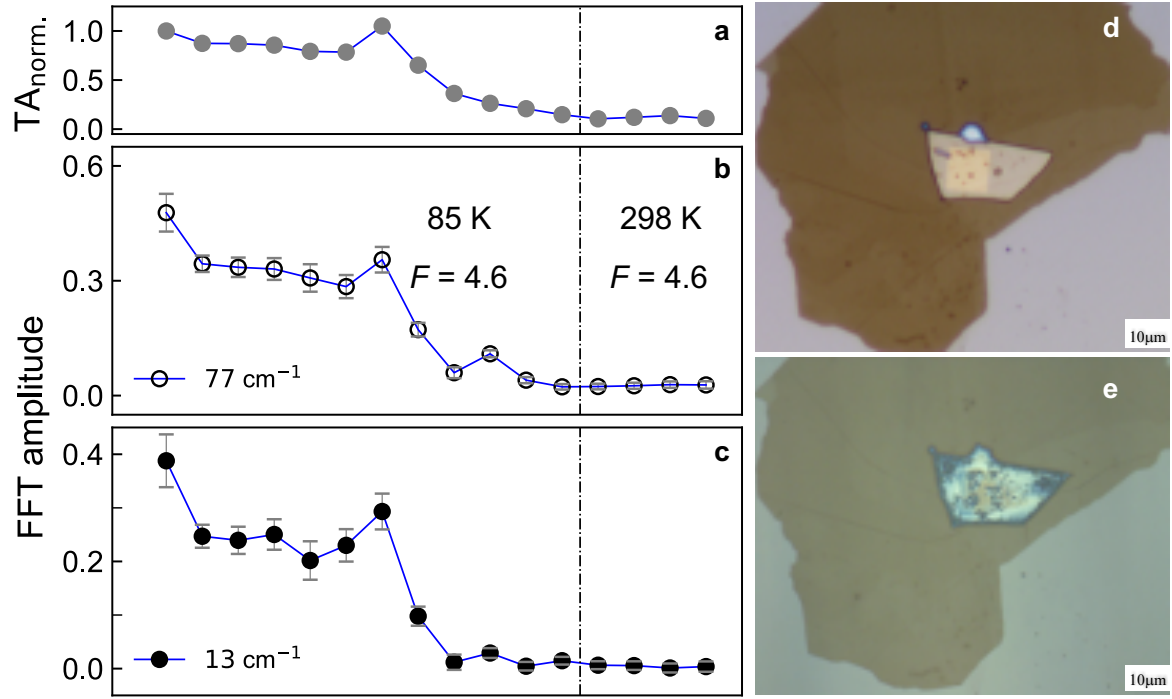

**Fig. S10: Laser induced changes as a function of the cumulative number of laser irradiation pulses.** From left to right the number of cumulative laser shots is 0, 10, 20, 30, 40, 50, 100, 200, 250, 300, 400, 900, 1000, 1100, 1300, 1900. **(a)** Recorded maximum magnitude of TA signal, normalized by its initial value. **(b)** FFT-amplitude of the 77 cm$^{-1}$ mode. **(c)** FFT-amplitude of the 13 cm$^{-1}$ mode. The temperatures and incident fluences employed in irradiation exposures are indicated in the figure. All tr-bb-TA measurements were carried out at 85 K, $F \approx 0.4$ mJ·cm$^{-2}$. **(d)** Microscope image of the pristine 1$T$'-MoTe$_2$ film. **(e)** Microscope image of the 1$T$'-MoTe$_2$ film after laser damage (1900 shots).

## VI. Laser Irradiation of Pristine 1T'-MoTe₂ Phase

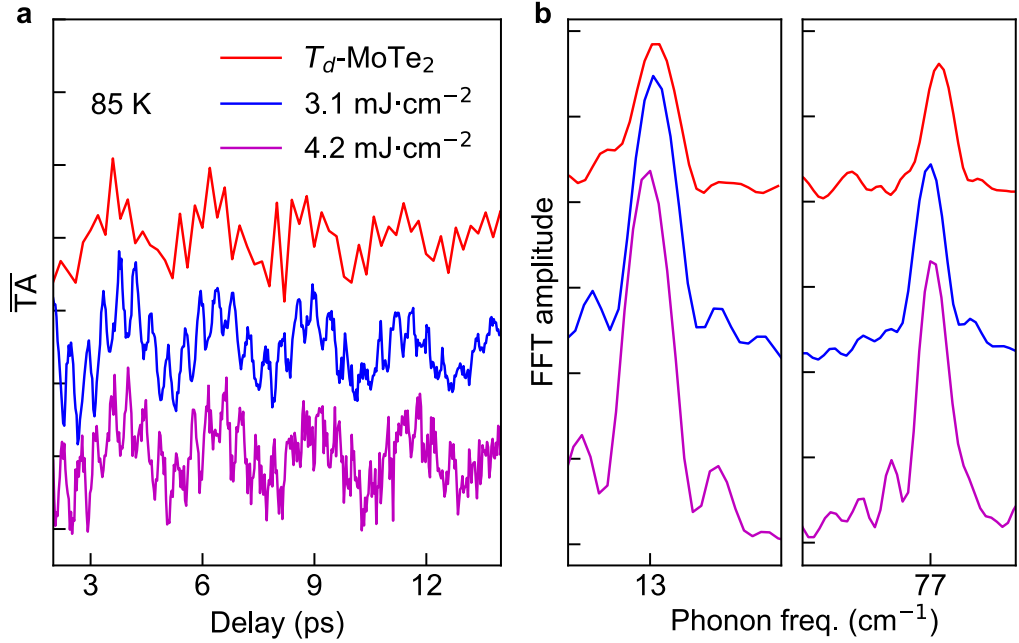

**Fig. S11: Laser irradiation test of a 30-nm thick 1T'-MoTe₂ flake. (a)** Persistence of the $^1A_1$ phonon mode upon increasing $F$ (inset). The sample was irradiated at 298 K and cooled down to 85 K for measuring phonon dynamics. The time traces have been averaged within $E \sim 1.38 - 1.55$ eV to improve the SNR. Even at the highest irradiation level of $F = 4.2$ mJ·cm$^{-2}$ the trace clearly shows the $^1A_1$ phonon mode characteristic of the $T_d$-phase. The first irradiation process was conducted for 5 minutes at 10 Hz (3000 pulses with $F = 3.1$ mJ·cm$^{-2}$). The second irradiation process was conducted for 10 minutes at 10 Hz (6000 pulses with $F = 4.2$ mJ·cm$^{-2}$). Tr-bb-TA measurements were performed at 85 K and $F = 0.5$ mJ·cm$^{-2}$. $E_{pump} = 2.4$ eV. The red trace corresponds to the pristine sample, and it was recorded with a coarser time step of 150 fs instead of 30 fs. **(b)** Frequency spectra obtained from each trace via FFT. The observed persistence of the coherent $^1A_1 \sim 13$ cm$^{-1}$ shear phonon mode in the $T_d$-state indicates that $T^*$ can only be formed if the initial state is the $T_d$-phase.

## VII. Transport Measurements as a Function of Temperature

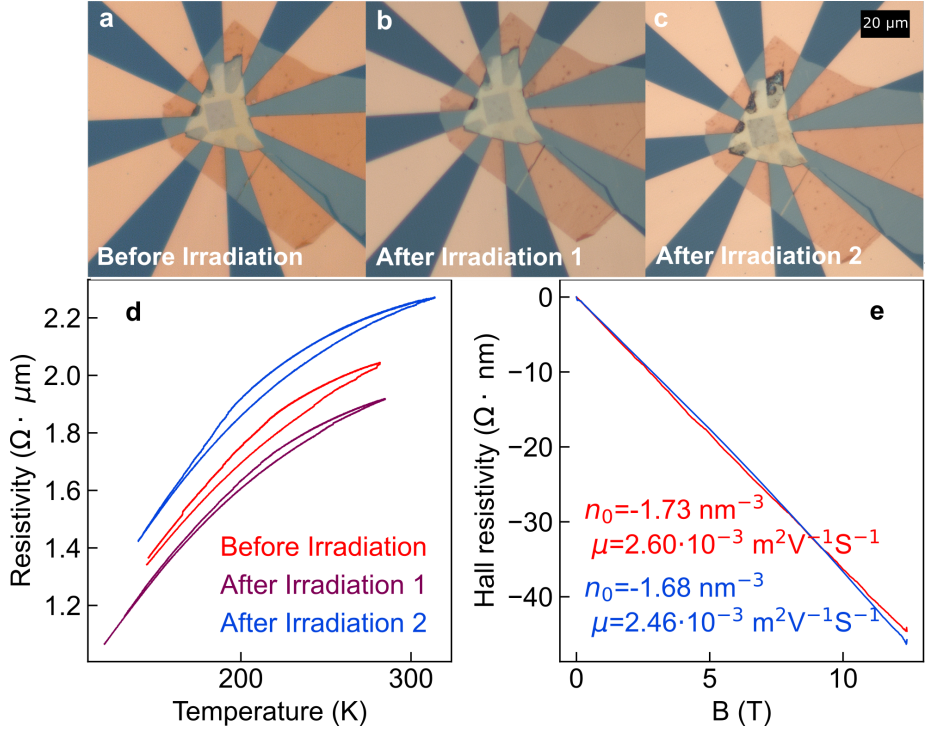

**Fig. S12: Resistivity measurements in a 30-nm thick MoTe$_2$ film supported by a Si frame with a central 10 μm x 10 μm Si$_3$N$_4$ window and Au electrodes in a star configuration. (a)** Microscope image of a pristine flake. **(b)** Same sample as in panel **a** but after laser irradiation using 2.5 mJ·cm$^{-2}$ at 85 K (5 minutes at 10 Hz). **(c)** Same sample as in panel **b** but after laser irradiation using 6.5 mJ·cm$^{-2}$ at 85 K (5 minutes at 10 Hz). It is noteworthy to mention that small bubbles of inert gas trapped during sample preparation appear as small darker spots on the sample. In addition, some laser-induced damage of the sample on top of the Au electrodes becomes visible. This damage seems not to largely affect transport measurements. **(d)** Resistivity measurements were carried out in a four-wire scheme after each irradiation. **(e)** Carrier density ($n_0$) and mobility ($μ$) were determined by measuring the Hall resistivity at 150 K before (red) and after the third irradiation (blue). Red traces in panels **d** and **e** correspond to the pristine sample. The purple and blue traces correspond to the same flake after a first laser irradiation (irrad. 1) at $T = 85$ K with 3000 pulses and $F = 2.5$ mJ·cm$^{-2}$ and after a second irradiation (irrad. 2) at $T = 85$ K, 3000 pulses and $F = 6.5$ mJ·cm$^{-2}$, respectively.

## References

[1] H.L. Kiwia and E.F. Westrum Jr, *J. Chem. Thermodyn.* **7**, 683 (1975).

[2] A.R. Beal, J.C. Knights, and W.Y. Liang, *J. Phys. C: Solid State Phys.* **5**, 3540 (1972).

[3] S. Song, D.H. Keum, S. Cho, D. Perello, Y. Kim, and Y.H. Lee, *Nano Letters* **16**, 188 (2016).